# The Radio Lifetime of Supernova Remnants and the Distribution of Pulsar Velocities at Birth

D. A. Frail and W. M. Goss

National Radio Astronomy Observatory, P.O. Box 0, Socorro, NM 87801

and

J. B. Z. Whiteoak

School of Physics, University of Sydney, Sydney, NSW 2006, Australia

## ABSTRACT

We have made VLA images of the fields around three young pulsars which have resulted in the discovery of two new supernova remnants and confirmation of a third. We argue that, in at least two cases and perhaps the third, the pulsars are physically associated with these supernova remnants. A review of all known young pulsars shows that the majority are associated with supernova remnants. We show that the typical density of the interstellar medium into which the supernova remnants are evolving has a density of 0.2 cm$^{-3}$ instead of the low value of 0.01 cm$^{-3}$ which had been calculated from other studies, and results in a considerably longer radio lifetime for supernova remnants. Both the morphology of the supernova remnants and the location of the pulsars imply that most of these young pulsars are born with large transverse velocities ($\sim$500 km s$^{-1}$). This high velocity mean in the distribution of pulsar velocities appears to be a general property of the pulsar population at birth, not seen in proper motion studies, due to selection effects. We explore the implications of this result as it relates to the origin of these velocities and the galactic distribution of pulsars. High velocity pulsars can escape their supernova remnant in a very short timescale, comparable to the lifetime of the remnant and may even play a role in extending the observable radio lifetime of the remnant. A significant fraction will be capable of escaping the disk of the Galaxy, producing an extended halo population.

*Subject headings:* (ISM:) supernova remnants, (stars:) pulsars: PSR 1643−43, PSR 1706−44, PSR 1727−33

## 1. INTRODUCTION

Young pulsars offer direct insight into some of the central questions concerning the origin and evolution of neutron stars. Statistical studies of the pulsar population as a whole have established beyond a doubt that the progenitors of pulsars are massive stars and that their birthplaces are the OB associations, dense gas and spiral arms from which such stars arise. This evidence includes the galactic distribution of pulsars (Lyne, Manchester & Taylor 1985), their kinematics (Lyne, Anderson & Salter 1982), and their birthrates (Narayan & Ostriker 1990). However, despite the insight that has been gained from these indirect methods, statistical analysis can never be a substitute for direct observational evidence. Lacking any recent supernovae in our our Galaxy, we turn to young pulsars and their associated supernova remnants.

With a median age of $4\times10^6$ years (Taylor, Manchester & Lyne 1993), the average pulsar is presently well away from its birthplace. Four million years is close to the lifetime of an OB association and it is well in excess of the age expected for a supernova remnant surrounding the pulsar. Thus it is only by studying the environments around *young* pulsars that we can hope to learn more about the birthplace of neutron stars. Here we define young pulsars, somewhat arbitrarily, as those pulsars with characteristic ages $\tau_c$ <60,000 years (=$P/2\dot{P}$, where P is the pulsar period and $\dot{P}$ is its period derivative). Through the study of young pulsars and their supernova remnants we can gain information about the periods, magnetic fields, and velocities of pulsars at their birth. The supernova remnant provides an independent age and distance estimate for the system and acts as a probe of the ambient gas (density, filling factor, pressure, etc) (Shull, Fesen & Saken 1989). The interaction between the energetic wind of a pulsar and the surrounding medium can give rise to pulsar wind nebula – a unique diagnostic of the pulsar magnetosphere (Wang, Li & Begelman 1993, Bietenholz, Frail & Hankins 1991).

The difficulty in pursuing such studies has been the small number of pulsar/supernova remnant associations (Helfand & Becker 1984). In the last five years major advances have been made on two fronts. High frequency radio surveys have penetrated farther into the plane of our Galaxy (Clifton et al. 1992, Johnston et al. 1992) and renewed radio imaging efforts (e.g. Caswell et al. 1992, Frail & Kulkarni 1991, Kassim & Weiler 1990a) have turned up several new associations. There are perhaps as many as 13 supernova remnants in association with young pulsars, and another 2 in "adolescent" pulsars (60 kyrs< $\tau_c$ < 110 kyrs). Recently, Johnston et al. (1992), in a 1.5 GHz survey of the southern Galactic plane, found five additional young pulsars. This brings the total number of young pulsars to 18 – still less than 3% of the total population of known pulsars.

Advances in low frequency radio imaging have also been needed in order to make deep, wide-field images toward these young pulsars. In particular, the development of powerful new reduction and imaging software (Cornwell 1993, Cornwell, Uson & Haddad 1992, Cornwell & Perley 1992) at the Very Large Array (VLA) has prompted the present study to do observations at 90-cm (0.3 GHz) and 20-cm (1.4 GHz) toward three of these pulsars in hopes of finding their putative supernova remnant. The observations are detailed in §2 and we discuss our results in §3. The implications of our detections are discussed in §4.

## 2. OBSERVATIONS

The observations were made with the VLA in the CnB (1993 Jun. 2) and DnC (1993 Sept. 14) hybrid array configurations. The fields toward three pulsars were observed (PSR 1643−43, PSR 1706−44, and PSR 1727−33) at 90 cm (0.3 GHz) and 20 cm (1.4 GHz).

The 20-cm observations were carried out in standard continuum mode. Two 50 MHz channels were measured in each hand of circular polarization at 1385.1 MHz and 1464.9 MHz. At 90-cm two separate IF bands, each measuring both hands of circular polarization and each with a bandwidth of 3.125 MHz, were used, at 327.5 MHz and 331.5 MHz. Each IF bandwidth was divided into 32 channels and the interference-free channels were selected and summed to form a final continuum database for further processing. All data reduction and calibration was done following standard practice in use at the VLA. The total integration time on each source was approximately 1 hour at both 20 cm and 90 cm, spaced over a full range of hour angle.

The data from each array were co-added to form a single dataset for each pulsar, which was then self-calibrated and CLEAN'ed to form the final radio image. The 90-cm data had to undergo specialized processing. The problems encountered when imaging in the galactic plane at 90-cm are detailed in Frail, Kassim & Weiler (1994). In short, the confusion produced by the numerous sources in the field must be dealt with by using a wide-field imaging algorithm developed by Cornwell & Perley (1992). Large $6.5° \times 6.5°$ fields result from this process and the rms noise in the galactic plane is typically reduced to 3-5 mJy, a factor of 10 below that achieved by standard techniques.

## 3. RESULTS

### 3.1. PSR 1643−43

PSR 1643−43 has a period P of 232 msec and a period derivative $\dot{P}$ of $113 \times 10^{-15}$ s s$^{-1}$, implying a surface dipole field B of $5.1 \times 10^{12}$ G (B=$10^{12}$ G$\sqrt{P \dot{P}_{-15}}$, where P is in seconds and $\dot{P}_{-15}$ is in units of $10^{-15}$ s s$^{-1}$) and a rotational energy loss rate $\dot{E}$ ($\dot{E} \propto P^{-3} \dot{P}$) of $3.6 \times 10^{35}$ erg s$^{-1}$ (Johnston & Manchester 1994). Its characteristic age $\tau_c$ is 32.6 kyrs ($\tau_c = P/2\dot{P}$) and its dispersion measure based distance is 6.9 kpc (Taylor & Cordes 1993).

Our 90-cm and 20-cm images toward PSR 1643−43 are shown in Figs. 1 and 2, respectively. In both images the pulsar appears to be embedded in an extended radio source. The diameter of this source is $22' \times 16'$, with its major axis lying in an east-west direction. In addition to this elliptical shell there is a "spur" of emission near the southwestern edge which runs north-south and blends into the extended background at 90 cm. The total flux density of this source (after making a correction for the primary beam of the antennas at both frequencies) is 1.25±0.05 Jy at 20 cm and 3.0±0.1 Jy at 90 cm. The angular structure represented in the images at both frequencies are similar. There is no evidence that we have missed a significant fraction of the total flux density of this source. Thus the source is definitely non-thermal with a spectral index $\alpha = -0.6$ ($S_\nu \propto \nu^\alpha$), a value typical of shell-type supernova remnants. The non-thermal character of this source is supported by the lack of extended emission in the 60 $\mu$m and 100 $\mu$m IRAS images of this region. Its 1 GHz surface brightness $\Sigma$ is $6.5 \times 10^{-22}$ W m$^{-2}$ Hz$^{-1}$, for which the $\Sigma$-D relation gives distances for the remnant between 8.3-9.7 kpc (Milne 1979, Clark & Caswell 1976, Allakhverdiyev et al. 1983). The geometric center of the remnant is at $\alpha(1950)=16^h\ 44^m$, $\delta(1950)=-43°\ 42'$, to which we assign the galactic source name of G 341.2+0.9.

PSR 1643−43 is located 8′ nearly due west of the geometric center of G 341.2+0.9 with a position angle of $-80°$. There is no point source at the timing position for PSR 1643−43 above a flux density of 0.5 mJy in the 20-cm image (Fig. 2). There is a 1.5 mJy point source just 40.3″ north of the timing position (Johnston & Manchester 1994). Since this is close to the cataloged flux density of 1.4 mJy we suggest that this point source at $\alpha(1950)=16^h\ 43^m\ 16.9^s\ (\pm 0.1^s)$, $\delta(1950)=-43°\ 40'\ 35''\ (\pm 2'')$ is actually the pulsar and that the timing position is in error. The derivation of timing positions for young pulsars is problematic because of rotational instabilities, which include both stochastic timing noise and secular "glitches" (Lyne 1992). If PSR 1643−43 were born at the geometric center of the remnant and traveled for $\tau_c = 32.6$ kyrs to its present location, it would have to be moving about 15 mas yr$^{-1}$, a transverse velocity 475 (d/6.9 kpc) km s$^{-1}$.

In addition to the shell-type emission surrounding the pulsar there is diffuse emission in the immediate vicinity of PSR 1643−43. This includes a 4′ nebulosity just east of the pulsar which is joined to the pulsar by a "bridge" of emission. The appearance of this structure is consistent with the implied westward motion of the pulsar. Ram pressure due to the pulsar's motion $\rho V^2_{PSR}=5.3 \times 10^{-9}\ n$ ergs cm$^{-3}$, where $n$ is the gas density, initially confines the pulsar wind and forces an otherwise spherically symmetric wind to flow out in a direction opposite to the pulsar's proper motion. As the wind expands adiabatically away from the pulsar, the dominant pressure term becomes the internal thermal pressure ($\simeq 10^{-10}$ erg cm$^{-3}$) in the evacuated cavity left by the blast wave as it sweeps up and heats the ISM. The energy requirements to power this emission are small. Assuming that the spectral index $\alpha$ is a constant $-0.6$ from $10^7$ Hz to $10^{11}$ Hz the radio luminosity of the *entire* remnant is only $1.4 \times 10^{33}$ erg s$^{-1}$, a tiny fraction of the $\dot{E}$ PSR 1643−43. Thus we suggest that the pulsar is powering a small plerionic remnant inside the shell of G 341.2+0.9 which expands away from the direction of motion of PSR 1643−43.

There is also a distinct brightening of G 341.2+0.9 on the side closest to PSR 1643−43. The radio morphology of G 341.2+0.9 is strongly reminiscent of G 5.4−1.2 (Frail, Kassim & Weiler 1994, Becker & Helfand 1985) and CTB 80 (Angerhofer et al. 1981). Both of these supernova remnants contain high velocity pulsars which have moved a significant distance from their birthplace and are currently exerting considerable influence on the evolution of the decelerating shell. In the case of G 5.4−1.2 and CTB 80 there is a noticeable brightening of the radio emission on the side of the remnant closest to the pulsar (Frail & Kulkarni 1991, Strom 1987). This behavior is explained in the "rejuvenation" model put forth by Shull, Fesen & Saken 1989. The energy loss from the spindown of young pulsars comes out of the magnetosphere in the form of a relativistic flow of particles and magnetic field. As the pulsar catches up and eventually overtakes the decelerating supernova remnant the relativistic particles and field from the pulsar will mix with the weak shock-driven emission from the shell, causing a local brightening of the remnant. Thus the extended radio emission surrounding PSR 1643−43 is powered by the pulsar, and there has been some early injection of these pulsar-generated particles into the shell, causing a brightening of western side of G 341.2+0.9.

This interaction may also explain the "spur" of emission to the south of G 341.2+0.9. A similar steep spectrum ridge is seen in CTB 80 and it coincides with a "breakout" of the remnant into a lower density region of the ISM (Fesen, Shull & Saken 1988). The compressed magnetic field lines from the shell channel the relativistic particles from the pulsar along this ridge, producing its unusual morphology.

The morphological evidence – the $4'$ nebulosity east of the pulsar, and the brightening of the shell closest to the pulsar – are strong indicators that a real physically association exists between this supernova remnant and the pulsar. PSR 1643−43 was born at the geometric center of G 341.2+0.9 and traveled to its present location creating the features that we observe. The dispersion measure-based distance to PSR 1643−43 of 6.9 kpc, and the $\Sigma$-D distance to G 341.2+0.9 of 9.0 kpc are broadly consistent with an association. Given the criticisms which have been leveled at the $\Sigma$-D relation (Green 1991), in particular selection effects which bias the relation against low surface brightness remnants like G 341.2+0.9, we adopt a distance to the system of 6.9 kpc.

### 3.2. PSR 1706−44

PSR 1706−44 has a period of 102 msec and a period derivative of $93 \times 10^{-15}$ s s$^{-1}$, implying a surface dipole field of $3.1 \times 10^{12}$ G and a rotational energy loss rate $\dot{E}$ of $3.4 \times 10^{36}$ erg s$^{-1}$ (Johnston & Manchester 1994). Its characteristic age is 17.5 kyrs and its dispersion measure based distance is 1.8 kpc (Taylor & Cordes 1993). Interest in PSR 1706−44 is high as it is one of only a 4 young pulsars for which pulsed emission has been detected at X-ray and gamma ray energies (Becker et al. 1992, Thompson et al. 1992). All three of the other young pulsars, PSR 0531+21 (Crab), PSR 0833−45 (Vela) and PSR 1509−58, are associated with known supernova remnants. Recently, McAdam, Osborne & Parkinson (1993) proposed that a faint arc of radio emission, on which PSR 1706−44 is superimposed, was in fact a supernova remnant and that the two objects were physically associated. Our deep VLA observations at 90-cm and 20-cm better delineate the full extent of the remnant and confirms an earlier speculation by McAdam et al. (1993).

Fig. 3. is a subsection from our wide field 90-cm image, showing the region around

PSR 1706−44. The brighter features in the 843 MHz image of McAdam et al. (1993) are plainly visible but in addition faint emission is seen further to north. The dominant radio emission comes from a well-defined limb-brightened shell. Weak, diffuse emission with a mean intensity of 7 mJy beam$^{-1}$ fills the region, both inside and outside the shell. A 32 arcmin diameter circle with a geometric center at approximately $\alpha(1950)=17^h\ 04^m\ 25^s$, $\delta(1950)=-44°\ 11.5'$ fits the data. Thus the galactic source name of the remnant is G 343.1−2.3. The bright point source near the center of the remnant has a steep spectral index $\alpha = -0.9$ between 327 MHz and 843 MHz, suggesting that it is likely a compact extragalactic source seen in projection against G 343.1−2.3. The total flux of G 343.1−2.3 at 90-cm (after subtracting point sources) is 10.6 Jy. McAdam et al. (1993) concluded that G 343.1−2.3 is a non-thermal source on the basis of single-dish flux densities and the lack of IR emission. The radio surface brightness at 1 GHz (assuming a typical spectral index of −0.5) is $9\times10^{-22}$ W m$^{-2}$ Hz−1 sr$^{-1}$. The $\Sigma$-D distance derived for G 343.1−2.3 by McAdam et al. (1993) is 3.0 kpc, and our slightly smaller measurement of $\Sigma$ implies an even larger distance. The dispersion measure distance of 1.8 kpc is favored for the pulsar because of constraints imposed by its gamma ray emission. There are theoretical predictions and observational confirmation that the efficiency of converting rotational spindown energy of the pulsar $\dot{E}$ into gamma ray emission increases inversely as $\dot{E}$ (Ulmer 1994, Dermer & Sturner 1994). At a distance of 1.8 kpc PSR 1706−44 fits on this relation but at 3.0 kpc the required efficiency is close to 10%, a value more typical of Geminga, which has an $\dot{E}$ two orders of magnitude smaller. Thus we favor a closer distance to the pulsar, and if the association is to hold up we require that G 343.1−2.3 be a nearby, under-luminous supernova remnant.

The 20-cm image in Fig. 4. gives a better view of the region immediately around the pulsar. Only a limited region is visible on this image because of the 30′ half power diameter of the VLA at this frequency. The pulsar is located 23′ south-east of the geometric center, on the outside edge of remnant. At the location of the pulsar the geometric center is at a position angle of −52 degrees. To have traveled to the edge of G 343.1−2.3 in 17.5 kyrs PSR 1706−44 the pulsar would have to be moving at 80 mas yr$^{-1}$, implying a transverse velocity of 670 (d/1.8 kpc) km s$^{-1}$.

The pulsar is an 11 mJy point source at 20-cm which is 19% polarized. McAdam et al. (1993) noted that the flux density for PSR 1706−44 varied by 25% or more over the course of several observations. Our flux density differs by a similar amount from the cataloged flux (8.6 mJy). This is presumably from interstellar scattering and is another point in favor in arguing for a nearer 1.8 kpc distance to PSR 1706−44. Polarized flux was not detected from any of the extended emission surrounding the pulsar. A fit for the position of the pulsar gives $\alpha(1950)=17^h\ 06^m\ 5.11^s\ (\pm0.03^s)$, $\delta(1950)=-44°\ 25'\ 23.5''\ (\pm0.3'')$. This is in good agreement with the interferometric position given by McAdam et al. (1993) but it disagrees with the position derived from radio timing observations by 15.7″ (Johnston & Manchester 1994).

Establishing a real physical association between G 343.1−2.3 and PSR 1706−44 is difficult. Given the uncertainties with the $\Sigma$-D relation the distance discrepancy discussed above may be overcome. However, unlike PSR 1643−43 there is no obvious sign that the pulsar is interacting with the remnant despite the fact that its $\dot{\mathrm{E}}$ is 10 times larger. G 343.1−2.3 is brightest to the northwest, the side closest to the galactic plane (Fig. 3). Closer to the pulsar the emission to the west is uniformly bright, while in the east the emission is weak. The pulsar lies near the end of the partial arc that defines the remnant. There is some indication in the 20-cm image in Fig. 4. that the relativistic wind from PSR 1706−44 is powering a small radio nebula. PSR 1706−44 is surrounded by a 4′ diameter "halo" and a 20′ "tail" falling off to the west. Such morphology is reminiscent of the cometary nebula surrounding PSR 1757−23 outside G 5.4−1.2 (Frail & Kulkarni 1991). If the pulsar originated from the geometric center of the G 343.1−2.3 then the ram pressure from the ISM is expected to produce a cometary nebula, with a tail which points back to the geometric center. The standoff point $r_w$ where pressure equilibrium is reached between the ram pressure of the ISM $\rho V_{\mathrm{PSR}}^2$ and the pulsar's wind $\dot{\mathrm{E}}/4\pi r_w^2 c$ is at $0.01/\sqrt{n_o}$ or $1.1''$ for $n_o=1.0$ cm$^{-3}$. None of these predictions are consistent with the halo/tail morphology that we see around PSR 1706−44. The halo is too large and the tail does not point back to the center of the remnant 23′ to the north-west. For this reason we consider an association between PSR 1706−44 and G 343.1−2.3 as unlikely but cannot rule it out. An alternative explanation is that G 343.1−2.3 is a background object. The tail is just part of the shell emission of G 343.1−2.3, while the halo is a plerion powered by PSR 1706−44 a low-velocity pulsar. Proper motion measurements should settle this issue.

### 3.3. PSR 1727−33

PSR 1727−33 has a period of 139 msec and a period derivative of $85.0\times10^{-15}$ s s$^{-1}$, implying a surface dipole field of $3.4\times10^{12}$ G and a rotational energy loss rate $\dot{\mathrm{E}}$ of $1.2\times10^{36}$ erg s$^{-1}$ (Johnston & Manchester 1994). Its characteristic age is 26.0 kyrs and its dispersion measure based distance is 4.2 kpc (Taylor & Cordes 1993).

Our 90-cm and 20-cm images toward this pulsar are shown in Fig. 5 and 6. The relatively high noise levels in these images are due to a bright HII region NGC 6357 (Caswell & Haynes 1987) located 1° from the phase center, which limits the final dynamic achieved. The pulsar is visible in the 20-cm image as a 3.7 mJy point source, close to its cataloged flux density of 3.0 mJy, and it is highly polarized ($> 46\%$). A fit for the position of the pulsar gives $\alpha(1950)=17^h 27^m 14.26^s$ ($\pm0.07^s$), $\delta(1950)=-33°\ 48'\ 19.4''$ ($\pm0.7''$). As with the two previous pulsars, this interferometric position disagrees (primarily in declination) with the position derived from radio timing observations by $7.6''$ (Johnston & Manchester 1994).

PSR 1727−33 is located in a complex region with several nearby extended sources. We have used our data along with a 60 $\mu$m IRAS image and a 843 MHz image from the MOST telescope (Whiteoak & Green) to identify these extended sources. The two bright sources to the north-east and south-east of PSR 1727−33 are the HII regions G 354.486+0.085 and G 354.2−0.054 with $\alpha = +0.10$ and $\alpha = +0.13$, respectively. The arc-like feature to the north-west of PSR 1727−33 is non-thermal ($\alpha = -0.24$) and is possibly a portion of a previously unknown supernova remnant shell. If the pulsar is part of a complex including the nearby HII region G 354.2−0.054 then we can determine a kinematic distance on the basis of H109$\alpha$ and H110$\alpha$ observations by Caswell and Haynes (1987). Assuming a flat rotation curve (Fich, Blitz & Stark 1989), rotation constants $R_o$ = 8.5 kpc and $\theta_o$=220 km s$^{-1}$, and an uncertainty of ±7 km s$^{-1}$, to allow for deviations from pure circular rotation, the $V_{LSR}$=−33 km s$^{-1}$ for the recombination lines gives a near distance between 4.7 and 5.6 kpc. Given the uncertainties involved we consider the agreement between this and the dispersion measure distance of 4.2 kpc to be good. Since the pulsar does not seem to be found toward a shell-type remnant we cannot derive a $\Sigma$-D distance.

In the low resolution 90-cm image the pulsar is found at the peak of extended emission. This suggests that at least some of this radio emission may be related to the pulsar, although a clear signature of a shell-like remnant as in the previous two pulsars is lacking. A clearer picture emerges in the 20 cm images. Most of the extended emission to the south-east of the pulsar is resolved out in these higher resolution observations, and the pulsar appears to lie at the base of a long, linear feature. This emission "contrail" appears to originate from the pulsar, moving originally north-west with a "kink" at $\delta$=−33° 45′ 20″ thereafter it bends in a more northerly direction where it can be see at least 10′ from the pulsar and perhaps at far away as 17′. We name this extended source G 354.1+0.1. The emission from G 354.1+0.1 is non-thermal, although the exact spectral index is hard to determine because of there is some evidence of missing extended flux at 20 cm. Excluding the pulsar, $\alpha = -0.17$ for the entire source but is steeper $\alpha = -0.6$ in the immediate vicinity of the pulsar, before the "kink". G 354.1+0.1 is certainly not a typical shell-type or pulsar-powered SNR, but its close match in distance with the pulsar, and the location of the pulsar at the tip of the contrail, both suggest that a real physical association exists between the two.

G 354.1+0.1 has the axisymmetric morphology of a rare class of non-thermal radio sources which include G 357.7−0.1 (Shaver et al. 1985, Becker & Helfand 1985), G 5.4−1.2 (Frail & Kulkarni 1991), and G 359.2−0.8 (Yusef-Zadeh & Bally 1987). Although it was once believed that the radio emission from these sources was powered by an accreting binary (Helfand & Becker 1985), it is now more likely that it arises from the spindown energy of a young, high velocity pulsar (Predehl & Kulkarni 1994). G 354.1+0.1 most resembles G 357.7−0.1. Both sources have long, linear emission and a compact radio source at the tip of the source and along the symmetry axis. The radio luminosity $L_R$ of G 354.1+0.1, found by integrating the $\alpha = -0.17$ from $10^7$ Hz to $10^{10}$ Hz is $1.5 \times 10^{32}$ erg s$^{-1}$. The minimum energy in field and particles, calculated in the usual way (Pacholczyk 1970), is $1.5 \times 10^{47}$ ergs, with a equipartition magnetic field $B_e$ of 16 $\mu$G. Since $L_R << \dot{E}$ for PSR 1727−33, it would have little difficulty in powering the nebular emission. The implied motion of the pulsar is in a direction opposite the contrail. At radio wavelengths for $B_e$=16 $\mu$G, the synchrotron lifetime $\tau_{syn}$ is of order $10^7$ yrs. Since $\tau_c << \tau_{syn}$, we can get a reasonable velocity of the pulsar by dividing the source size by the pulsar age $\tau_c$. The angular extent of this linear feature gives a size of 12 pc (d/4.2 kpc), in good agreement with the sizes of the other members of this class. To have traveled this distance in the 26 kyrs requires a transverse velocity of 460 km s$^{-1}$(d/4.2 kpc), or a proper motion of 23 mas yr$^{-1}$.

## 4. DISCUSSION

### 4.1. The Mean Lifetimes of Supernova Remnants

Shortly after the discovery of pulsars nearly 25 years ago only the Crab and Vela pulsars had an associated supernova remnant. This situation remained relatively static until the last decade, during which there was a slow and steady increase in the number of known associations. This list has grown to the point where there are as many as (perhaps) 17 associations. Where once such associations with young pulsars were the exception rather than the rule, the opposite now appears to be case. We summarize our findings in Table 1 and compare them with the properties of other young pulsars. For each pulsar we list its period, period derivative, a distance, and characteristic age ($\tau_c$ =P/2$\dot{\rm P}$), magnetic field (B=$10^{12}\,G\sqrt{\rm P\dot{P}}$), and spin down luminosity ($\dot{\rm E}\propto{\rm P}^{-3}\,\dot{\rm P}$) taken from Taylor, Manchester & Lyne (1993). If a supernova remnant is known to be associated with the pulsar the remnant's name is given. Various derived quantities are also given in the last columns, which are defined later in the text. Some of these associations are controversial and so in the final column we include references which discuss the evidence of the association (both for and against).

Some of the associations in Table 1 are not as secure as others and much effort needs to be made to test whether these associations are true. However, there are three pulsars in Table 1 without *any* associated supernova remnant: PSR 1046−58, PSR 1737−30, and PSR 1823−13. The fields around PSR 1737−30 and PSR 1823−13, as well as PSR 1800−21, were imaged by Braun, Goss & Lyne (1989) with the VLA at 20 cm. They failed to detect extended emission around these pulsars. This apparent lack of shell-type or (pulsar-powered) plerionic supernova remnants around these young pulsars lead Braun, Goss & Lyne (1989) to conclude that the mean lifetimes of supernova remnants were of order 20,000 yrs or less. This was taken as evidence that the progenitors of pulsars exploded in low density environments ($n_o \simeq 0.01$ cm$^{-3}$), resulting in the rapid expansion and dissipation of the remnant, which faded beyond detectability after $\simeq 10^4$ yrs (Bhattacharya 1990). However, as we now know that supernova remnants have been found around all young pulsars to which adequate radio imaging observations have been made. We must re-examine the conclusions of Braun, Goss & Lyne (1989).

For a supernova remnant in the Sedov phase its radius is given by R$_s(t)$ = $12.4(E_{51}/n_o)^{1/5}(t_s/10^4)^{2/5}$ pc, where E$_{51}$ is the blast energy in units of $10^{51}$ ergs and $t_s$ is the remnant's age in years (Shull, Fesen & Saken 1989). For older remnants a more accurate formula exists by Cioffi, McKee & Bertschinger (1988). If we assume that the spindown of the pulsar is dominated by magnetic dipole radiation and that its initial period at birth is small (i.e. P$_i$ <<P), then the pulsar's characteristic age is a good measure of the remnant's age $\tau_c \simeq t_s$ (Kulkarni et al. 1988). This allows useful constraints to be put on the ratio $n_o/E_{51}$. For each supernova remnant in Table 1 which has a well-defined shell, we computed the ratio $n_o/E_{51}$. We used angular diameters from Green (1991) when they were available, and values from the table references when they were not. The pulsar distances for calculating $R_s$ were taken from Table 1. The majority of the distances are based on dispersion measures (Taylor & Cordes 1993), but if more direct measurements were available (e.g. HI absorption for W 28) these were used instead.

The results of these calculations are given in column (7) of Table 1. For $t_s$ <2 kyrs the values of $n_o/E_{51}$ should not be trusted as these remnants are still likely in the free expansion phase. X-ray estimates of $E_{51}$ from historical supernova remnants give $E_{51}$=0.2 (Smith 1988). Optical, radio and X-ray data of galactic supernova remnants with known distances give a range of $E_{51}$ from 0.4 to 2.0 (Berkhuijsen 1988), while extragalactic supernovae give $E_{51}$=0.5 to 2.0 (Wheeler & Filippenko 1994, Arnett et al. 1989). Thus we can assume $E_{51} \simeq 1$ and calculate $n_o$ directly. The average density is approximately 0.2 cm$^{-3}$ with a few objects showing large excursions in either direction. While this calculation of $n_o$ is over-simplified and depends sensitively on distance, there is clearly little evidence that these supernova are expanding into medium with density of order 0.01 cm$^{-3}$. To derive such low values would require that we systematically underestimate the pulsar distances by almost a factor of 2. The Taylor and Cordes (1993) model has some difficulties with estimating distances in the inner Galaxy where many of our pulsars are found. However, the errors arise from approximating the electron density in the inner Galaxy by an average value, thereby underestimating the contribution of discrete HII regions, which decrease these dispersion measure distances, not increase them. The low values for G 308.8−0.1 and G 57.1+1.7 may be due to the large dispersion measure-based distances that were used. Use of the surface brightness-diameter relation ($\Sigma$-D) gives smaller distances to each remnant of 6.9 and 4.5 kpc, respectively (Caswell et al. 1992, Routledge & Vaneldik 1988). The large value of $n_o$ for W 28 is expected, as it is known to be expanding into a dense molecular cloud (Frail, Goss and Slysh 1994).

We conclude from our data that there is no reason to support the assertion by Braun et al. (1989) that the density of the ISM around supernova remnants is of order 0.01 cm$^{-3}$. On spatial scales of a few parsecs to tens of parsecs the majority of these supernova remnants are propagating into media whose densities are typical of the warm neutral and warm ionized phases of the interstellar medium (Kulkarni & Heiles 1988). Furthermore, it follows that the mean lifetime of radio supernova remnants is $> 60,000$ years and not 20,000 years as Braun et al. (1989) concluded because they failed to detect supernova remnants around young pulsars. In §4.3 we will examine the role that pulsars may have in prolonging the lifetimes of shell-type supernova remnants.

The real reason for the non-detections by Braun, Goss & Lyne (1989) is more mundane. The extended radio emission that we detected around PSR 1643−43, PSR 1706−44, and PSR 1727−33 are from large ($> 15'$), low surface brightness remnants, not easily imaged with the limited field of view of the VLA at 20 cm. The 1 GHz surface brightnesses of the three supernova remnants observed in this paper is $<10^{-21}$ W m$^{-2}$ Hz sr$^{-1}$. This is low, although Braun et al. (1989) should still have been capable of detecting these particular supernova remnants at 20 cm. Our surface brightnesses and angular sizes are similar to the 11 new supernova remnants found by Taylor,

Wallace & Goss (1992) in a 327 MHz survey of the plane with the Westerbork Synthesis Radio Telescope. They found that interferometric observations at 327 MHz were highly successful in discovering new supernova remnants that higher frequency and/or single dish surveys had failed to detect. It is particularly telling that later VLA images of PSR 1800−21 at 90 cm, one of the 3 pulsars surveyed by Braun et al. (1989), reveal a 45′ supernova remnant W 30 (Frail, Kassim & Weiler 1994, Kassim & Weiler 1990a). Braun et al. (1989) failed to detect this source at 20 cm because of the much reduced sensitivity of the VLA to faint structures on large angular scales ($> 15'$). This holds the promise that new VLA imaging at 90 cm of the fields around PSR 1737−30 and PSR 1823−13 may reveal the faint remnants associated with these pulsars. Radio imaging of a one square degree region around PSR 1046−58 with the Molonglo Observatory Synthesis Telescope (MOST) at 843 MHz reveals no associated extended emission at the 2 mJy/beam level with a 43" beam (Whiteoak & Green 1994).

### 4.2. The Distribution of Pulsar Velocities at Birth

Gunn and Ostriker (1970) were first to show that the spatial distribution of pulsars was consistent with a population born close to the plane ($|z| < 70$ pc) and with high enough velocities ($> 100$ km s$^{-1}$) that over time they were carried to their large observed scale heights ($h_z \simeq 400$ pc). This central tenet of the genesis of pulsars has stood up well to various tests, including a much increased pulsar sample (Taylor, Manchester & Lyne 1993) and pulsar proper motion measurements (Lyne, Anderson & Salter 1982). Proper motion surveys in particular have provided direct evidence that pulsars are high velocity objects. Interferometric measurements (Harrison, Lyne & Anderson 1993, Bailes et al. 1990) give $<V_{PSR}> \simeq 200$ km s$^{-1}$, while scintillation measurements give $<V_{PSR}> \simeq 100$ km s$^{-1}$ (Cordes 1986).

Establishing the underlying distribution of pulsar velocities at birth from these measurements is difficult. The selection effects in proper motion surveys and their bias against high velocity pulsars are well understood (Cordes 1986, Helfand & Tademaru 1977). In short, for a given age, large velocity pulsars occupy a much larger volume than small velocity pulsars. Thus in these largely distance-limited samples (i.e. bright, nearby pulsars), the high velocity pulsars are grossly under-represented in the sample and therefore proper motion surveys may not be sampling the true distribution of pulsar velocities. This point is illustrated by Harrison et al. (1993) where they show that the younger and the more distant pulsars have a higher proportion of large $<V_{PSR}>$ values than the older and/or nearby pulsars. The average velocity between these different datasets differs by a factor of three.

By directly measuring the velocities of young pulsars we can hope to recover the true distribution of pulsar velocities. Unfortunately proper motion measurements have only been made for the Vela and Crab pulsars (Bailes et al. 1989, Wyckoff & Murray 1977). For the remaining pulsars in Table 1 we use an indirect method first suggested by Shull et al. (1989). For the majority of the associations in Table 1 the pulsar is offset substantially from the center of the supernova remnant (e.g. Fig 2. and 3.). We can quantify this through the parameter $\beta=\theta_p/\theta_s$ where $\theta_p$ is the angular displacement of the pulsar from the geometric center of the remnant, $\theta_s$ is the angular radius of the supernova remnant. For values of $\beta > 1.0$ the pulsar is outside the remnant (e.g. W 28, G 5.4−1.2).

If the pulsar was actually born at the geometric center of the supernova remnant then the values of $\beta$ in Table 1 signify large proper motions and hence high values of V$_{\rm PSR}$. For each pulsar in Table 1 we calculated its *implied* transverse velocity V$_{\rm PSR}$ by dividing its transverse displacement from the geometric center of the remnant by its age $\tau_c$. These transverse velocity values are found in column (11) of Table 1. For the youngest pulsars ($\tau_c <5$ kyrs) small offsets of the pulsar from the center of the remnant, due perhaps to slight asymmetries in the expansion of the remnant, get magnified by the young age, resulting in unreliable velocity values. The high value of V$_{\rm PSR}$ for PSR 1509−58 likely arises from this effect. Likewise for PSR 0540−69, we derive V$_{\rm PSR}$=260 km s$^{-1}$, but Manchester, Staveley-Smith & Kesteven (1993) derive an implied V$_{\rm PSR}$=1200 km s$^{-1}$ with slightly different assumptions.

The velocities we derive from this method can be compared to the distribution of velocities from the compilation of proper motions by Harrison et al. (1993). The 15 pulsars in Table 1 have a very different distribution than those of Harrison et al. (see Fig. 7). Twelve of the 15 young pulsars have velocities above the mean of 217 km s$^{-1}$ for the Harrison et al. (1993) sample. The mean velocity of our sample is 990 km s$^{-1}$ with a median of 480 km s$^{-1}$. Questionable associations, yielding high transverse velocities, will skew the mean to higher values, whereas the median is less affected. If we discard three of the more controversial associations in Table 1 with some of the highest velocities (i.e. PSR 1509−58, PSR 1610−50 and PSR 1930+22), the mean drops to 550 km s$^{-1}$, while the median is relatively unchanged at 460 km s$^{-1}$. These large V$_{\rm PSR}$ values have been noted before in discussions of individual associations (e.g. Kassim & Weiler 1990a, Manchester et al. 1991), but it was Caraveo (1993) who first suggested that this may be a general property of young pulsars as a whole.

The validity of these implied velocities rests on the assumption that the geometric center of the remnant is the blast center of the supernova event. As a supernova remnant expands and interacts with an inhomogeneous ISM the birthplace of the pulsar may become difficult to define. Radio brightness gradients towards the plane, barrel-shaped remnants, and breakout morphologies attest to the importance of the ISM in shaping supernova remnants (Caswell 1988). Despite the fact that they evolve in a complex environment, the majority of supernova remnants (including those in Table 1) do retain circular symmetry (Dubner et al. 1993). These well-defined circular morphologies impose a severe constraint on the degree to which asymmetric expansion of a supernova remnant can cause the blast and geometric centers to diverge. In a recent example, Frail et al. (1994) found that they could shift the blast center away from the geometric center of G 5.4−1.2 by no more than 25% without producing a significant deviation in the observed shape of the remnant. This approach is flawed if the true shape of the SNR is not known. This is illustrated by PSR 0833−45 in the Vela XYZ radio nebula. If the pulsar were born at the geometric center, as defined by the cm-radio data, then its implied velocity would be 800 km s$^{-1}$. Interferometric measurements of its proper motion yield a velocity of only 120 km s$^{-1}$ (Bailes et al. 1989). This discrepancy originates from the poorly defined shape of the remnant, leading to an incorrect estimate of the geometric center. Low frequency radio observations (Dwarakanath 1991) and X-ray observations (Aschenbach 1992) of the Vela remnant show faint emission to the west, revealing a larger and more circular remnant than had been previously suspected. The new geometric center of the remnant is now consistent with the blast center inferred from the pulsar's proper motion. Thus the velocities derived from pulsar displacements in supernova remnants with poorly defined shapes (i.e W 30, G 57.1+1.7, Kes 32) are suspect. Distance uncertainties on average will contribute another 25% to the uncertainty on $V_{PSR}$. However, even allowing for some asymmetric expansion of the supernova remnant into the ISM and the distance uncertainties, it is difficult to see how the velocities for the remaining pulsars in Table 1 with well-defined remnants are in error by a factor of two. With this larger sample we conclude, like Caraveo (1993), that the *inferred* velocities of young pulsars are greatly in excess of those pulsar velocities *measured* by proper motion studies. Given the known bias in proper motion studies this result implies that there exists a significant high velocity component in the true distribution of pulsar velocities.

After this paper had been submitted for review we became aware of a preprint by Lyne & Lorimer (1994) in which they reassessed the velocities derived from proper motion studies, using the newer Taylor and Cordes (1993) distance models and correcting the sample for the selection effect discussed earlier. They found a mean pulsar birth velocity of 450±90 km s$^{-1}$, in line with the value derived here from young pulsars in supernova remnants.

### 4.2.1. *The Distribution of Pulsars in the Galaxy*

The implications of the existence of this increased mean in the velocity distribution is far-reaching; it has an impact on the mechanisms that give rise to pulsar velocities at birth, and it affects how they interact with their surroundings and what the final distribution of pulsars in the Galaxy will be.

The derived scale height of pulsars ($h_p = 400$ pc) (Lyne, Manchester & Taylor 1985) does not appear at first to be consistent with a large population of $V_{PSR} > 500$ km s$^{-1}$ pulsars. Such a population would quickly leave the plane, increasing the scale height well above that which is observed unless there were significant luminosity and/or magnetic field evolution. However, Frei, Huang & Paczyński (1992) and Paczyński (1990) showed that the scale height of pulsars is dominated by low velocity objects and that the contribution by high velocity pulsars (which spend only a tiny fraction of their time in the disk) is small. Furthermore, in order to derive the *true* vertical distribution of pulsars in the Galaxy assumptions need to be made concerning instrumental selection effects and the choice of a distance scale. Lyne et al. (1985) have described how improper accounting for instrumental selection effects have affected prior conclusions about the pulsar scale height. Bhattacharya & Verbunt (1991) point out that because of the finite scale height of the ionized gas layer, from which the majority of distances are derived, pulsars at large scale heights are improperly accounted for in these statistical studies. A proper accounting of the various effects is beyond the scope of this paper but future statistical studies should recognize that the velocity distribution may be very different to what has been previously assumed.

High velocity pulsars with a significant z-component in their velocity vector ($V_z \sim 400$ km s$^{-1}$) may escape the disk, forming a halo population of old neutron stars. This extended halo population has been postulated to exist for many years in order to explain gamma-ray bursts as a galactic phenomenon (Shklovski & Mitrofanov 1985, Paczyński 1990, Hartmann, Epstein & Woosley 1990, Li & Dermer 1992). A substantial fraction of pulsars in Table 1 ($\epsilon \sim 0.35$) have such z-velocities. In $10^8$ years when the neutron star has ceased to be a pulsar it is 40 kpc out of the plane.

Paczyński (1993), however, has criticized these halo models, pointing out that the high velocity pulsars have no special property that (period, magnetic field, etc.) that distinguishes them from their counterparts in the disk. This is only partially correct as the magnetic fields of these pulsars are larger on average than the mean for all pulsars (see §4.2.2). However, none of the fields of these young pulsars are in the range of the $10^{14}$-$10^{15}$ G fields evoked in some models of bursters (e.g. Duncan & Thompson 1992). Other objections raised by Paczyński and others about the halo models still remain. Regardless of whether these pulsars are the progenitors of gamma ray bursts, we find it notable that a significant fraction of the pulsar population may have escaped the disk and are not being properly accounted for in space density and birthrate estimates.

### 4.2.2. Possible Mechanisms for Producing High Velocity Pulsars

Dewey and Cordes (1987) showed that the binary hypothesis for the origin of pulsar velocities has difficulty in producing a population of very high velocity pulsars. Some sort of velocity "kick" is required at the time of the supernova explosion even to match the observed distributions. Several mechanisms have been proposed, all of which involve some sort of explosion asymmetry. Early ideas dealt with asymmetric mass ejection of the stellar envelope (Shklovskii 1969) and a similar asymmetry was been suggested for the neutrino burst (Woosley 1987). More recent efforts suggest that Rayleigh-Taylor instabilities or large scale convective motions of the material in the proto-neutron star, just prior to the supernova explosion, is a potential source of large recoil velocities (Burrows & Fryxell 1992, Herant, Benz & Colgate 1992). However, Colgate & Leonard (1994) have suggested that the highest velocity kicks from convective processes come from tidally locked, close binary systems where the rotation period of the neutron star is $>> 1$ sec. Lower modes of rotation result in larger asymmetries during convection. None of the high velocity pulsars in Table 1 have such high periods and thus would not be expected to get a large kick from the Colgate and Leonard mechanism.

There is a suggestion in Table 1 that the high velocity pulsars have correspondingly high dipole magnetic fields. The magnetic field of these pulsars exceeds the mean log B of 12.51 (Stollman 1987), although not by large factors. The maximum field is only a order of magnitude greater than this, and the more typical values exceed the mean by only 2 or 3. These high B-field values place them at the upper end of the velocity-magnetic field correlation (Bailes 1989 and references therein). The two pulsars with the largest magnetic fields (PSR 1509−58 abd PSR 1610−50) also have the largest inferred transverse velocities. This correlation, whose origin has long been a puzzle since it was first noted by Helfand & Tademaru (1977), suggests that the magnetic field is an active agent in producing pulsar velocities (Li & Dermer 1992). Alternatively, one can view this as a selection effect, since young pulsars are more luminous than "average" pulsars and their luminosity depends on magnetic field to some positive power, usually assumed to lie between 1 and 2 (Frail & Moffett 1993). However, attempts to explain the correlation as a selection effect have been unsuccessful (Dewey & Cordes 1987, Stollman & van den Heuvel 1986).

Anderson & Lyne (1983) suggested that an asymmetry in the supernova explosion was responsible for the pulsar velocity, and that it could be influenced by the magnetic field in some unspecified way. Radhakrishnan (1985) has criticized this suggestion, pointing out that the magnetic energy density ($B^2/8\pi$) over the volume of the neutron star is a negligible fraction of the total energy released in a supernova explosion ($\sim 0.1\,\mathrm{Mc}^2$) and therefore its effect must be minimal. The fractional isotropy required to provide the momentum impulse to the neutron star is small, only 1-2% of the total energy (Chugai 1984) for $V_{PSR}$=500 km s$^{-1}$, but it is much larger than that which can be provided by a global B-field of order $10^{12}$ G. Chugai (1984) showed that in the presence of a magnetic field the neutrinos, which carry off the bulk of the energy in a supernova explosion, could be emitted asymmetrically (i.e. parallel to the B-field). Preliminary calculations showed that the velocity recoil imparted to the newly formed neutron star was small unless the magnetic field was $10^{14}$ G or more. A variety of other mechanisms were proposed by Duncan & Thompson (1992) but here again the majority were only important in strong fields ($10^{14}$-$10^{15}$ G). Local B-field enhancements could result in the necessary asymmetries, but we know of no working neutron star models that explicitly account for such affects.

### 4.2.3. Interacting Composites

In the taxonomy of supernova remnants it was once thought that there existed a clear distinction between the shell-type remnants, which are powered by a shock wave as it interacts with the swept-up gas from the ISM, and the plerions, which are powered by the relativistic particles and magnetic field that flow out from a young, energetic pulsar (Reynolds 1988). Composite remnants, which show both the characteristics of shells and plerions, are also known, but the two evolve separately from each other (Helfand & Becker 1987, Helfand & Becker 1985). The discovery of the pulsar in CTB 80 changed this, it blurred the distinction between the classes and resulted in an entirely new class of "interacting composite" remnants.

CTB 80 is an old ($\sim 10^5$ yrs) supernova remnant in the radiative phase whose peculiar morphology is the result of its interaction with PSR 1951+32. Shull, Fesen & Saken (1989) have explained in detail how pulsars are able to "rejuvenate" aging remnants through such interactions. Again, however, these authors calculated that interacting composites would be relatively rare because for typical pulsar velocities of 100-200 km s$^{-1}$ the time needed for the pulsar to catch up to a shell was of order a few $\times 10^5$ yrs. Furthermore, PSR 1951+32 is special. Its small dipole field (log B=11.68) and small $\dot{P}$ give it a spindown luminosity 100$\times$ larger than other pulsars its age.

Interacting composites neither appear to be as rare as initially supposed nor do they necessarily occur as late in the evolution of a supernova remnant as once thought. In addition to CTB 80 and the new objects discussed in this paper, there are several other candidates for interacting composites in the literature, each with varying degrees of certainty. These include a number of objects mentioned by Shull et al. (1989) like G 5.4-1.2, W 28 and G 57.1+1.7 (Frail & Kulkarni 1991, Frail, Kassim & Weiler 1994, Frail, Kulkarni & Vasisht 1993, Routledge & Vaneldik 1988), as well as G 114.3+0.3 (Kulkarni et al. 1993), G308.8−0.1 (Kaspi et al. 1992), and MSH 15−52 (Caswell, Milne & Wellington 1981). At radio wavelengths the tell tale signs of an interacting composite are a brightening of the supernova remnant closest to the pulsar, a bow shock around the pulsar, a flat spectral index close to the pulsar which steepens further away, and a high degree of polarization. This large number of young, interacting composites is a direct consequence of high pulsar velocities and is more indirect evidence that the high inferred velocities in Table 1 are correct.

With V$_{PSR}$=500 km s$^{-1}$ a pulsar catches up to its supernova remnant in only 40,000-70,000 years (Shull, Fesen & Saken 1989). On these much shorter timescales the supernova remnant is likely still in the Sedov phase and thus the pulsar can escape the boundaries of its supernova remnant while it can be a prominent multi-wavelength source. Future pulsar searches near supernova remnants should recognize this fact. Such pulsars will act as a "fountain of youth" injecting fresh relativistic particles and field into the compressed shell of the aging remnant. Thus the ambient density will not be the sole factor influencing the mean lifetimes of SNRs.

## 5. CONCLUSIONS

Deep imaging has been made in the fields toward three young pulsars with the VLA at 20 cm and 90 cm. In all three cases the pulsars are found nearby or within extended radio emission, which we believe to be supernova remnants. It is argued on the basis of like distances and morphological evidence that there is a true physical association between the objects in at least two of these three cases.

With these discoveries we bring the total number of possible pulsar-supernova remnant associations to 17. Fifteen of the 18 youngest pulsars ($\tau_c$ <60,000 yrs) have some sort of an associated remnant. Taken as a whole, the young pulsars and their associated remnants suggest that the mean age of radio supernova remnants is at least 60,000 years and that the mean velocity of pulsars may be as large as 500 km s$^{-1}$. The implications for the existence of this high velocity component in the velocity distribution is far-reaching; it has an impact on the mechanisms that give rise to pulsar velocities at birth, and it affects how they interact with their surroundings and what the final distribution of pulsars in the Galaxy will be.

Some of the associations in Table 1 are controversial and much work needs to be done to test whether they are true. Radio and X-ray observations offer the best opportunity to get independent distances, shock velocities, ages, blast energies, etc. for these supernova remnants. Proper motion measurements of the pulsars are the best test of both the veracity of the associations and our prediction of high pulsar velocities ($\sim$500 km s$^{-1}$). The former is tested by the direction of the velocity vector (which should be away from the geometric center of the remnant), while the later is tested by its magnitude. The pulsar proper motions, implied by their offsets from the geometric centers of their remnants, predict values between 10-100 mas yr$^{-1}$. With suitable nearby reference sources it should be possible to measure the proper motions for the majority of the pulsars in Table 1 in the next few years with either the VLA or the VLBA.

Instead of concentrating on the centers of supernova remnants, future searches for young pulsars might be well advised to look for signs of a recent interaction of the pulsar with the remnant (brightened arcs, high polarization, flat spectral index, etc.) and concentrate on these regions.

Acknowledgements: DAF thanks Shri Kulkarni and Steve Thorsett for reading of an early version of the manuscript and acknowledges useful conversations with S. Colgate and J. Cordes. DAF thanks D. Helfand for calling attention to the fact that the inferred velocities of young pulsars were all on the high end of the measured velocity distribution. This research has made use of the SIMBAD database, operated at CDS, Strasbourg, France. The Very Large Array (VLA) and the Very Long Baseline Array (VLBA) are operated by Associated Universities, Inc. under cooperative agreement with the National Science Foundation.

Fig. 1.— A 90-cm radio continuum image of the field around PSR 1643−43. Uniform weighting was used and the synthesized beamsize is $116'' \times 54''$, PA=18°. The rms noise in the image is 3.3 mJy beam$^{-1}$. Contour intervals are 5, 15, 25, 35, 45, 55, 65, 75 mJy beam$^{-1}$. The position of PSR 1643−43 is indicated by the cross.

Fig. 2.— A 20-cm radio continuum image of the field around PSR 1643−43. Uniform weighting was used and the synthesized beamsize is $25'' \times 21''$, PA=0°. Contour intervals are 0.5, 0.75, 1.0, 1.5, 2.0, 2.5, 3.0, 3.5, 4.0 mJy beam$^{-1}$. The rms noise is 0.25 mJy beam$^{-1}$. A primary beam correction has not been applied. The pulsar is the 1.5 mJy point source, located at $\alpha = 16^h\ 43^m\ 16.9^s$, $\delta = -43°\ 40'\ 35''$, and indicated by the arrows.

Fig. 3.— A 90-cm radio continuum image of the field around PSR 1706−44. Uniform weighting was used and the synthesized beamsize is $111'' \times 56''$, PA=12°. The rms noise in the image is 3.0 mJy beam$^{-1}$. Contour intervals are 5, 15, 25, 35, 45, 55 mJy beam$^{-1}$. The position of PSR 1706−44 is indicated by the cross. The horizontal stripes visible at $\delta = -44°\ 40'$, and $\delta = -43°\ 47.5'$ are slight imperfections which result from the joining of the data to form the final wide field image.

Fig. 4.— A 20-cm radio continuum image of the field around PSR 1706−44. Uniform weighting was used and the synthesized beamsize is $27'' \times 22''$, PA=0°. Contour intervals are 0.25, 0.5, 0.75, 1.0, 1.25, 1.5 mJy beam$^{-1}$. The rms noise is 0.13 mJy beam$^{-1}$. A primary beam correction has not been applied. The pulsar is the 11 mJy point source indicated by the arrows. It is located at $\alpha = 17^h\ 06^m\ 5.11^s$, $\delta = -44°\ 25'\ 23.5''$, in the center of the 4' "halo" of extended emission.

Fig. 5.— A 90-cm radio continuum image of the field around PSR 1727−33. Uniform weighting was used and the synthesized beamsize is $99'' \times 47''$, PA=41°. The rms noise in the image is 3.5 mJy beam$^{-1}$. Contour intervals are -17.5 (dashed), 10, 17.5, 35, 52.5, 70, 105, 140, 280, 420 mJy beam$^{-1}$. The position of PSR 1706−44 is indicated by the cross.

Fig. 6.— A 20-cm radio continuum image of the field around PSR 1727−33. Uniform weighting was used and the synthesized beamsize is $21'' \times 8.8''$, PA=44°. Contour intervals are 0.8, 1.6, 2.4, 3.2, 4.0, 10, 30, 50, 100 mJy beam$^{-1}$. The rms noise is 0.4 mJy beam$^{-1}$. A primary beam correction has not been applied. The pulsar is the 3.7 mJy point source indicated by the arrows.

Fig. 7.— Histograms of pulsar velocities. The top plot shows the distribution of the 66 transverse velocities from Harrison et al. (1993), while the bottom plot shows the distribution of transverse velocities from Table 1.